\documentclass[useAMS,usenatbib,usegraphicx]{mn2e}

\newcommand{\kev}{keV}
\newcommand{\chandra}{\textit{Chandra}}
\newcommand{\integral}{\textit{INTEGRAL}}
\newcommand{\rxte}{\textit{RXTE}}
\newcommand{\heao}{\textit{HEAO-1}}
\newcommand{\bat}{\textit{Swift}-BAT}
\newcommand{\rosat}{\textit{ROSAT}}
\newcommand{\maxi}{\textit{MAXI}}

\newcommand{\xmm}{\textit{XMM-Newton}}
\newcommand{\bepposax}{\textit{BeppoSAX}}
\newcommand{\fe}{Fe~K$\alpha$}
\newcommand{\etal}{et al.}

\title[The average 0.5--200 keV spectrum of local AGNs and the 2--10~keV LF
  at $z \approx 0$]
  {The average 0.5--200~keV spectrum of local active galactic nuclei
    and a new determination of the 2--10~keV luminosity function at $\mathbf{z
  \approx 0}$}
\author[D.\ R.\ Ballantyne]
  {D.~R.~Ballantyne \thanks{david.ballantyne@physics.gatech.edu}\\Center for Relativistic Astrophysics, School of Physics, Georgia
  Institute of Technology, 837 State Street, Atlanta, GA 30332-0430, USA}

\date{Accepted 2013 October 28. Received 2013 October 25; in original form 2013 
August 19}

\pagerange{\pageref{firstpage}--\pageref{lastpage}}
\pubyear{2013}

\usepackage{times}

\begin{document}

\label{firstpage}

\maketitle

\begin{abstract}
The broadband X-ray spectra of active galactic nuclei (AGNs) contains
information about the nuclear environment from Schwarzschild radii
scales (where the primary power-law is generated in a corona) to
distances of $\sim 1$~pc (where the distant reflector may be
located). In addition, the average shape of the X-ray spectrum is an
important input into X-ray background synthesis models. Here,
local ($z \approx 0$) AGN luminosity functions (LFs) in five energy
bands are used as a low-resolution, luminosity-dependent X-ray
spectrometer in order to constrain the average AGN X-ray spectrum
between $0.5$ and $200$~keV. The $15$--$55$~keV LF measured by
\bat\ is assumed to be the best determination of the local LF, and
then a spectral model is varied to determine the best fit to the
$0.5$--$2$~keV, $2$--$10$~keV, $3$--$20$~keV and $14$--$195$~keV LFs. The spectral
model consists of a Gaussian distribution of power-laws with a mean
photon-index $\langle
\Gamma \rangle$ and cutoff energy $E_{\mathrm{cut}}$, as well as
contributions from distant and disc reflection. The reflection
strength is parameterised by varying the Fe abundance relative to
solar, $A_{\mathrm{Fe}}$, and requiring a specific \fe\ equivalent
width (EW). In this way, the presence of the X-ray Baldwin effect can be
tested. The spectral model that best fits the four LFs has $\langle \Gamma \rangle = 1.85 \pm 0.15$,
$E_{\mathrm{cut}}=270^{+170}_{-80}$~keV,
$A_{\mathrm{Fe}}=0.3^{+0.3}_{-0.15}$ (90\% C.L.). The sub-solar $A_{\mathrm{Fe}}$ is unlikely to be a true
measure of the gas-phase metallicity, but indicates the presence of
strong reflection given the assumed \fe\ EW. Indeed, parameterising the reflection strength with the $R$ parameter
gives $R=1.7^{+1.7}_{-0.85}$. There is moderate evidence for no X-ray
Baldwin effect. Accretion disc reflection is included in
the best fit
model, but it is relatively weak (broad iron
K$\alpha$ EW $< 100$~eV) and does not significantly affect any of the conclusions. A critical result of our procedure is
that the shape of the local $2$--$10$~keV LF measured by \heao\ and
\maxi\ is incompatible with the LFs measured in the hard X-rays by
\bat\ and \rxte. We therefore present a new determination of the local
$2$--$10$~keV LF that is consistent with all other energy bands, as
well as the de-evolved $2$--$10$~keV LF estimated from the \xmm\ Hard
Bright Survey. This
new LF should be used to revise current measurements of the evolving AGN
LF in the $2$--$10$~keV band. Finally, the suggested absence of the X-ray Baldwin
effect points to a possible origin for the
distant reflector in dusty gas not associated with the AGN obscuring
medium. This may be the same material that produces the compact
12$\mu$m source in local AGNs.
\end{abstract}


\section{Introduction}
\label{sect:intro}
The X-ray spectra of active galactic nuclei (AGNs) span nearly three
decades in energy and are comprised of many separate components: a
power-law with a high energy cutoff \citep*[e.g.,][]{mdp93,np94,rt00,zdz00,matt01,mol06,mol09,winter09,derosa12,rmr13,vas13}, reflection from both distant
material and the accretion disc \citep*[e.g.,][]{pou90,np94,tan95,fab02,nan97,nan07,ball10,dcp10,shu10,pat12,ricci13} and, in many cases, a
soft excess and/or a warm absorber \citep*[e.g.,][]{tp89,rey97,per02,blustin05,crummy06,scott11,scott12,tom13}. The variability properties of the X-ray emission
indicate that it is generated from very close to the black hole ($\la
 10$--$20$~$r_g$, where $r_g=GM/c^2$ is the gravitational radius of a black
hole with mass $M$; e.g., \citealt{gra92,mch06,uttley07,zog13}), a size scale so small that the physics of
the region can only be investigated by spectroscopy. For example, the
slope and cutoff energy of the primary power-law is related to the
temperature and optical depth of the Comptonizing corona \citep[e.g.,][]{zdz00,pet01,mol09}, and the disc
reflection features can probe the space-time of the central black hole
and the physical state of the underlying accretion flow \citep*[e.g.,][]{fab89,laor91,br06,br09,mill07,rf08,bmr11}. High quality broadband measurements of the X-ray AGNs are
therefore crucial to understanding the physics of the central
accretion disc \citep[e.g.,][]{ris13}.

In addition to the study of individual objects, studying how the
spectral properties of AGNs vary as a function of luminosity and/or
redshift may give potentially valuable insights into the evolution of AGNs
and the unified model. Current survey results indicate that the
X-ray coronal properties seem to be most dependent on the Eddington
ratio and not on the cosmic epoch \citep*{shem06,rym09,bright13}. However, distant
reflection, as measured from the equivalent width (EW) of the narrow
\fe\ line, has shown evidence for a inverse dependence on the X-ray
luminosity \citep[e.g.,][]{it93,bia07,shu10,ricci13}. This effect, if real \citep{shu12}, would then be providing
a clue on how the distant, potentially Compton-thick material around
AGNs is dependent on the central engine, with clear implications for
the AGN unified model \citep{ricci13}.

The broadband spectral shape of AGNs is also a key ingredient for
X-ray background (XRB) models \citep*[e.g.,][]{gch07,tre09,ball11,ay12}. It is now known that the XRB between $\approx 2$--$200$~keV is comprised of the
integrated observed emission of AGNs across all $z$ and X-ray
luminosities. Therefore, modeling the XRB requires knowledge of the
broadband spectral shape of AGNs, and how it may change with
luminosity and redshift. Uncertainty in the spectral shape, in
particular the high-energy cutoff and reflection strength, translates
directly into uncertainty in the Compton-thick AGN fractions that are
derived from fitting the peak of the XRB spectrum at $\approx
20$--$30$~keV \citep{ay12}. Currently, all XRB models 
assume a spectral shape with some authors accounting for the observed
distribution of photon indices and the apparent decline of reflection
strength with luminosity \citep{gch07,ball11}.  

Aside from a small number of bright sources observed with
\bepposax\ \citep{matt01}, the full energy range of AGN spectra has only been studied
by combining observations from different X-ray telescopes that can
only focus on a small part of the entire spectrum \citep*[e.g.,][]{derosa12,vmg13}. This
approach must deal with the cross-calibration of different instruments
and the fact that the various observations are often not performed
simultaneously, making the resulting spectrum susceptible to the
effects of spectral variability. As a result, while catalogues of the
spectral properties of hundreds of AGNs have been published \citep[e.g.,][]{bn11,scott11,rmr13,vas13}, these
results are often isolated from one another and a clear picture of the
broadband spectral properties of AGNs remains elusive. In this paper,
we present a novel approach to measure the average $0.5$--$200$~keV spectrum of
local AGNs by using the AGN luminosity function (LF) in four
energy bands as a low resolution, luminosity-dependent
spectrometer. The advantage of using the LFs in this way is that
instrumental and (in some cases) absorption effects have been
removed. The LFs are also constructed from observations of many AGNs
over timescales $\ga 1$~year, mitigating the effects of
variability. Thus, finding the spectral model that can simultaneously
fit the AGN LF from $0.5$ to $\approx 200$~keV will provide a
relatively unbiased view of the average $z \sim 0$ spectral shape.

In the next section, the methodology, LFs and spectral models that are used in the
calculation are described. The results of fitting the LFs are
discussed in Section~\ref{sect:res}, in particular the constraints on
the spectral model. Section~\ref{sect:discuss} compares the results to
previous works, and discusses the
implications on XRB modeling and the origin of the distant reflector. We
especially emphasize a new measurement of the local $2$--$10$~keV
LF. Finally, the conclusions of this study are summarized in
Sect.~\ref{sect:concl}. If applicable, all results in this paper make
use of the following cosmological parameters:
$H_0=70$~km~s$^{-1}$~Mpc$^{-1}$, $\Omega_{\Lambda}=0.7$, and
$\Omega_{m}=0.3$.

\section{Calculations}
\label{sect:calc} 

\subsection{Methodology}
\label{sub:method}
Given a broadband AGN X-ray spectral model, $L_{E}$, and a reference
LF measured in a specific energy band, $d\Phi/d \log
L_{\mathrm{ref}}$, then the LF can be predicted in any other band via
\begin{equation}
{d\Phi \over d\log L} = C(L) {d\Phi \over d\log L_{\mathrm{ref}}} {\Delta
  \log L_{\mathrm{ref}} \over \Delta \log L},
\label{eq:method}
\end{equation}
where $\log L$ is the luminosity in the new band calculated from
integrating $L_{E}$. The factor $C(L)$ can be used to convert LFs that
account for the entire AGN population to one that includes only a
subset. As is described below, an $C(L)$ factor is needed when
calculating the $0.5$--$2$~keV LF for unobscured AGNs. 

In this study, various forms of $L_E$ are considered and $d\Phi/d \log
L$ is calculated in four widely separated energy bands. Chi-squared
fitting is then used to determine the best fitting spectral model and
the error-bars on the associated parameters.

\subsection{Local X-ray Luminosity Functions}
\label{sub:xlfs}
The following five measurements of the local AGN LF are used to
constrain the average spectral shape of AGNs at $z \approx 0$.

\vspace{0.2cm}
\noindent$\bullet$\textbf{15--55~keV} \citep{ajello12}\\
This LF is derived from the 60 month \bat\ survey and consists of 428
AGNs with a median redshift of 0.029. As this LF is derived from the most recent and
least biased survey of Compton-thin AGNs in the local Universe, it is
used as the reference LF for this study. To ease the comparison with LFs derived from lower-energy
observations, we make use of the LF derived solely from Compton-thin
AGNs (\citealt{ajello12}; Table 5; third row). This LF is consistent
with other measurements of the local LF in similar energy ranges
\citep[e.g.,][]{saz08}.

\begin{figure}
\centerline{
\includegraphics[width=0.5\textwidth]{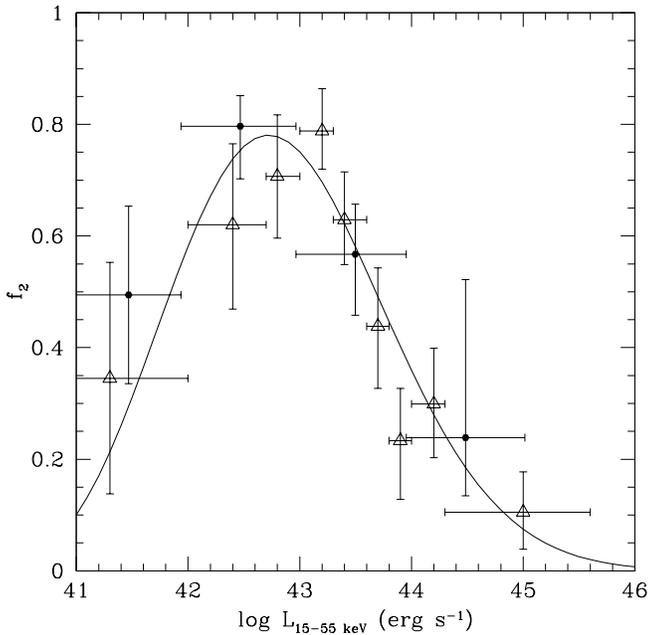}
}
\caption{The triangles plot the AGN obscured fraction assuming no
  Compton-thick sources, $f_2$, versus
  the 15--55~keV luminosity \citep{burlon11}. The solid
line shows the fit to these data described by eq.~\ref{eq:f2}. The
obscured fractions determined by \citet{bn11} are shown as the solid
points. These measurements were made as a function of 2--10~keV
luminosity, but, given the size of the error-bars, they were not
converted to 15--55~keV luminosities for this plot.}
\label{fig:f2vsL}
\end{figure}

\vspace{0.2cm}
\noindent$\bullet$\textbf{0.5--2~keV} \citep*{has05}\\
These authors published the local LF for Type 1 (i.e., non-obscured) AGNs
based on 205 AGNs with $0.015 < z \la 0.1$ from the \rosat\ Bright Survey. As
this LF is for only unobscured AGNs, and the reference $15$--$55$~keV
LF includes all Compton-thin AGN, a Type~1 fraction must be specified
in order to compare the predicted $0.5$--$2$~keV LF to the observed
data. Several surveys indicate that this fraction depends
strongly on AGN luminosity \citep[e.g.,][]{ueda03,simp05,della08,has08,gan09,burlon11,assef13}. The triangles in Fig.~\ref{fig:f2vsL} plot the Type 2 AGN
fraction, $f_2$, against the $15$--$55$~keV luminosity measured by
\citet{burlon11} from the 3-year \bat\ survey. Here, $f_2$ is defined
as the ratio of Compton-thin AGNs with $N_{\mathrm{H}} \geq
10^{22}$~cm$^{-2}$ to the total number of AGNs with $N_{\mathrm{H}} <
10^{24}$~cm$^{-2}$ (i.e., no Compton thick objects).  
The obscured fraction falls sharply with luminosity at both high
and low luminosity. The effect at high luminosity is often thought to
be a result of radiation pressure eroding the covering factor of the
dusty obscuring material \citep{law91}. The significance of the decrease of $f_2$ at low
luminosities is still tentative, but might indicate that a critical
luminosity is needed to produce and sustain a geometrically thick
obscuring zone \citep[e.g.,][]{eh09,ms13}. The solid line in Fig.~\ref{fig:f2vsL} plots
the following fit to the \citet{burlon11} data:
\begin{equation}
\label{eq:f2}
f_2=0.1 \left(\log L-40 \right)^{3} \exp(-0.32(\log
L-41)^2),
\end{equation}
where $L$ is the $15$--$55$~keV luminosity. Using the smallest of the two
error-bars on each $f_2$ data-point gives $\chi^2=6.7$ for 6 degrees of freedom
(dof). Fig.~\ref{fig:f2vsL} also shows that the $f_2$ fractions measured by
\citet{bn11} from a 12~$\mu$m selected sample (solid points) are in
good agreement with the \bat\ results and the best fit
model. Eq.~\ref{eq:f2} is used to calculate the Type~2 AGN fraction
when predicting the Type~1 $0.5$--$2$~keV LF (i.e., $C(L)=1-f_2$).

The \citet{has05} LF is based on AGNs identified as Type 1 via the
optical definition (i.e., broad Balmer emission lines), while the
\bat\ measurements and eq.~\ref{eq:f2} follow the typical X-ray
definition (i.e., obscuring line-of-sight column density
$N_{\mathrm{H}} < 10^{22}$~cm$^{-2}$). \citet{has08} argues that the
optical definition of an unobscured AGN corresponds to a lower value
of $N_{\mathrm{H}} \la 3\times 10^{21}$~cm$^{-2}$. Therefore, $C(L)$
computed using Eq.~\ref{eq:f2} must be corrected for
the percentage of AGNs with $3\times 10^{21}$~cm$^{-2} \la
N_{\mathrm{H}} < 10^{22}$~cm$^{-2}$. The $N_{\mathrm{H}}$ distribution
measured by \citet{burlon11} is used to make this small (only a 11\%
reduction) correction.

Finally, \citet{has05} corrects the $0.5$--$2$~keV luminosities of the
AGN sample for Galactic extinction, but not for any small amounts of
intrinsic absorption. Thus, all $0.5$--$2$~keV luminosities are
predicted for AGN spectra subject to obscuration with $\log
N_{\mathrm{H}}=20, 20.5$ and $21$, with weightings given
by the \citet{burlon11} $N_{\mathrm{H}}$ distribution.

\vspace{0.2cm}
\noindent$\bullet$\textbf{2--10~keV} \citep{shin06,ueda11}\\
The \citet{shin06} LF is constructed from 49 AGNs with $z < 0.4$
observed by \heao, while the \citet{ueda11} LF is produced from 37
sources with $0.002 < z < 0.2$ detected by \textit{MAXI}. Both
analyses corrected the $2$--$10$~keV luminosities for absorption and
are sensitive to Compton-thin AGNs. Interestingly, \citet{ueda11}
point out a disagreement between the shape of the derived
$2$--$10$~keV LF with the one measured by \bat\ in the $14$--$195$~keV
band. These authors suggest that if the spectral shape of AGNs steepens as
luminosity increases then the LFs can be made to agree. A similar
inconsistency between the \citet{shin06} LF and the $3$--$20$~keV LF
was discussed by \citet{della08} who showed that it could not be due
to absorption.

\vspace{0.2cm}
\noindent$\bullet$\textbf{3--20~keV} \citep{sr04}\\
This LF was constructed from 76 Compton-thin AGNs detected by the \rxte\ Slew
Survey. All but 6 sources are at $z < 0.1$. \citet{sr04} defined the
LF as the number density per observed luminosity interval, so the
\citet{burlon11} $N_{\mathrm{H}}$ absorption distribution was applied
to our spectral model prior to calculating the $3$--$20$~keV
luminosities when predicting this LF. The luminosities of the
\citet{sr04} LF are increased by 1.4 to account for an error in the
flux conversion \citep{saz08}.

\vspace{0.2cm}
\noindent$\bullet$\textbf{14--195~keV} \citep{tuell08}\\
This LF is based on 88 non-beamed AGNs detected in the
9-month \bat\ survey. The AGNs have a median redshift of $0.017$. Given
the high energy range of this survey, no absorption correction was
made to the luminosities. Although this LF and the $15$--$55$~keV LF
are not strictly independent, the latter LF is derived from a much
larger dataset in a different energy range by a different analysis
technique. Therefore, we consider these two LFs independent enough to
allow chi-squared fitting. 

\subsection{Model AGN Spectra}
\label{sub:model}
The AGN spectral model consists of a power-law with a photon-index
$\Gamma$ and a high energy cutoff, $E_{\mathrm{cut}}$, and one or two
reflection spectra, denoting distant and accretion disc
reflection. AGNs exhibit a range of $\Gamma$ that is roughly normally
distributed about some average with a dispersion of
$\sigma_{\Gamma}\approx 0.3$ \citep[e.g.,][]{bn11,scott11,derosa12,vas13}. The mean value of this distribution,
$\langle \Gamma \rangle$, is not observed to be correlated with the
X-ray luminosity \citep[e.g.,][]{bright13}. As the LFs are derived from observations of
several AGNs, the spectral model is constructed by Gaussian averaging 11
cutoff power-laws around a central $\langle \Gamma \rangle$ with
$\sigma_{\Gamma}=0.3$ and $d\Gamma=0.1$ (i.e., a $\langle \Gamma
\rangle=1.9$ model includes contributions from power-laws with
$\Gamma=1.4$--$2.4$; e.g., \citealt{gch07}). The distribution of $E_{\mathrm{cut}}$ remains
uncertain \citep[e.g.,][]{mol09,ricci11}, so, for simplicity, all AGNs are assumed to have the same
value of $E_{\mathrm{cut}}$. 

Distant reflection is modeled using the `pexmon' model \citep{nan07} available
in XSPEC \citep{arn96}. The advantage of this model is that it self-consistently
adds the \fe, Fe K$\beta$, Ni K$\alpha$ lines and the \fe\ Compton
shoulder to the reflection continuum accounting for how the strengths
of these features depend on $\Gamma$, the inclination angle and the
metal abundances. The origin of the distant reflector in AGNs is not
understood, but may be connected to the pc-scale obscuring material
(this is discussed further in Sect.~\ref{sub:distant}). In that case,
the inclination angle of the reflector could vary widely between
Type~1 and 2 AGNs. To account for this, an angle-averaged reflection
spectrum (running from 5~deg to 85~deg in steps of 5~deg) was
constructed for every $\Gamma$ and $E_{\mathrm{cut}}$. 

This angle-averaged reflection spectrum is then added to each
power-law prior to the performing the Gaussian average (the results do
not significantly change if the reflection spectra are Gaussian averaged and then
added to the averaged power-law spectrum). The $R$ parameter is traditionally used
to measure the reflection strength in an AGN spectrum. However, its
value only has physical meaning for the case of an isotropic source
above a disc. Alternatively, the reflection strength can be investigated by
specifying the \fe\ EW, and varying the iron abundance
($A_{\mathrm{Fe}}$, where $A_{\mathrm{Fe}}=1$ for Solar
abundances). In this manner, an interesting physical property (the Fe
abundance of the distant reflector) can be investigated. Spectral models are constructed with the
\fe\ EW set in one of two ways: a constant value of 70~eV, or
following the \chandra-derived X-ray Baldwin effect: $\log
\mathrm{EW}=1.64-0.11(\log L -44)$ \citep{shu10,shu12}. The Baldwin effect is
measured as a function of $2$--$10$~keV luminosity, but is nearly
identical for the $15$--$55$~keV luminosities employed here. The \citet{angr89}
abundance set is used throughout this study, and only
$A_{\mathrm{Fe}}$ is varied \footnote{The calculations
  were repeated with different abundance sets \citep[e.g.,][]{lodd03}, a constant
  inclination angle (30~deg or 60~deg), and with letting the abundance of all
  metals to be varied along with Fe. In all
  cases, the same best fit spectral model is found (within the
  uncertainties) and the qualitative results are unchanged.}.  

It is expected that most AGNs also exhibit a disc reflection component
including a relativistically broadened \fe\ line \citep[e.g.,][]{nan07}. This reflector
will contribute to the overall reflection spectrum and therefore may
impact the derived strength of the distant reflector. The disc reflection component is
investigated with a relativistically blurred `pexmon'
model. The `kerrconv' model \citep{br06} is used to blur the reflection
spectrum, assuming a black hole spin of $0.998$, and a disk that is
illuminated from $r_{\mathrm{ISCO}}$ to 400~$r_{\mathrm{ISCO}}$ with
an emissivity index of $3$ (ISCO=innermost stable circular
orbit). For simplicity, both the `kerrconv' and `pexmon' models have a
fixed inclination angle of $60$~deg. As with the distant reflector, the disc model is fixed to give a certain \fe\ EW (relative to the power-law;
the narrow \fe\ line is removed when adding the disc spectrum). Since the average broad \fe\ EW is observed
to be $\la 100$~eV \citep[e.g.,][]{nan07,dcp10,cha12,pat12}, disc reflectors
are considered with a \fe\ EW of either 54~eV
\citep{ball10} or 96~eV \citep{pat12}. The disc model has the same
$A_{\mathrm{Fe}}$, $\Gamma$ and $E_{cut}$ as the distant reflection
and power-law models. As is seen below, the addition of disc
reflection at this level has a very small impact on the fits to the LFs, so no
further investigation of this component is pursued.

In summary, the spectral model $L_{E}$ is constructed by adding
reflection spectra derived from the `pexmon' model to a cutoff
power-law. All three components have the same $\Gamma$ and
$E_{\mathrm{cut}}$. The strengths of the reflection components are set
by fixing the EW of the \fe\ lines to a specific value. The spectra are then Gaussian averaged around a mean $\langle \Gamma
\rangle$ and normalized to a specific $15$--$55$~keV luminosity to give the final spectral model. In this way, each $L_E$ is
determined by 3 parameters: $\langle \Gamma \rangle$,
$E_{\mathrm{cut}}$ and $A_{\mathrm{Fe}}$. The model is then used,
along with the $15$--$55$~keV LF, to predict the LFs in the
$0.5$--$2$~keV, $2$--$10$~keV, $3$--$20$~keV and $14$--$195$~keV
bands (Sect.~\ref{sub:method}). The parameters are varied with $\langle \Gamma
\rangle=1.7,\ldots,2$ (steps of $0.05$),
$E_{\mathrm{cut}}=50,\ldots,450$~keV (steps of $10$~keV), and
$A_{\mathrm{Fe}}=0.15,\ldots,2$ (steps of $0.05$), and the model that
best fits the observed LFs is determined by calculating the joint
$\chi^2$ (51 data points, 48 dof). A total of six different $L_E$ are
considered encompassing all the permutations of the different
reflection strengths (including the presence or absence of the disc component). In addition, a $L_{E}$ with no reflection
components is constructed to compare against the other, more
realistic models. All spectral models are defined from $0.1$~keV to
$900$~keV in 1000 logarithimically spaced steps. With this binning the
narrow \fe\ line has a width (at the base of the line) of 0.12~keV.

\section{Results}
\label{sect:res}
Here, we discuss the best fit average AGN spectral model at $z \approx
0$ derived from
fitting the local $0.5$--$2$~keV, $2$--$10$~keV, $3$--$20$~keV and
$14$--$195$~keV LFs. The uncertainty on each
parameter is calculated using the 90\% confidence level (C.L.) on the
parameter of interest (i.e., a $\Delta \chi^2=+2.71$ criterion). All $\chi^2$
are calculated using the smallest error-bar on each data-point.

It is interesting to first consider the result for a spectral model
constructed just from cutoff power-laws (i.e., no distant or disc
reflection). In this very idealized case, the best fit model is
$\langle \Gamma \rangle=1.7$ and $E_{\mathrm{cut}}=400$~keV with
$\chi^2/\mathrm{dof}=144/49=2.94$; a very poor fit. Adding distant
reflection to the model, with the \fe\ line following the Baldwin
effect, decreases $\chi^2$ by 53, with the addition of one additional
degree of freedom ($A_{\mathrm{Fe}}$). Therefore, the presence of
distant reflection in the average AGN spectrum is highly
significant (F-test probability$=3\times 10^{-6}$).

\begin{figure*}
\centerline{
\includegraphics[width=0.8\textwidth]{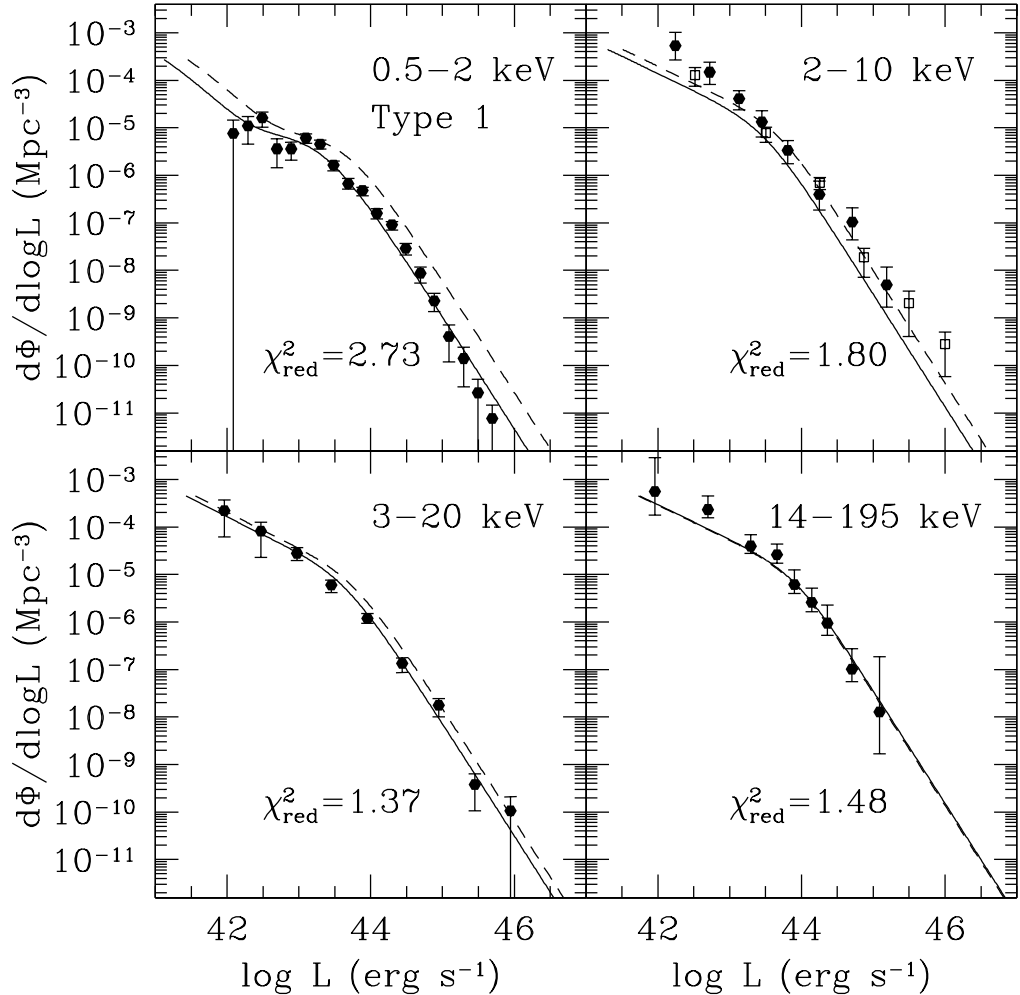}
}
\caption{The solid lines plot the X-ray LFs as predicted from the best
fit spectral model: no X-ray Baldwin effect, disc \fe\ EW$=54$~eV, $\langle \Gamma \rangle = 1.85 \pm 0.15$,
$E_{\mathrm{cut}}=270^{+170}_{-80}$~keV, and
$A_{\mathrm{Fe}}=0.3^{+0.3}_{-0.15}$ (90\% C.L.). The joint reduced
$\chi^2$ of this fit is $1.68$. The $\chi^2$ of the individual LFs are
indicated in the panels. The presence of disc reflection makes
only a minor improvement to the fit, but the addition of the Baldwin effect
results in $\Delta \chi^2=+9.4$. The data
points are from \citet{has05} ($0.5$--$2$~\kev), \citet{ueda11}
($2$--$10$~keV; solid points), \citet{shin06} ($2$--$10$~keV; open
points), \citet{sr04} ($3$--$20$~keV), and \citet{tuell08}
($14$--$195$~keV). In the $0.5$--$2$~keV band, the
Type~1 LF is predicted using the $f_2$--$L$ relation (eq.~\ref{eq:f2})
and the \citet{burlon11} $N_{\mathrm{H}}$ distribution to isolate the
fraction of AGNs with $\log N_{\mathrm{H}} \leq 21.5$ \citep{has08}. The
dashed lines show the best-fit 2--10~keV LF obtained by ignoring the
other 3 bands ($\chi^2_{\mathrm{red}} \mathrm{(2-10\ keV)}=0.7$;
$\chi^2_{\mathrm{red}} \mathrm{(Joint)}=33$; no Baldwin effect, disc
\fe\ EW$=54$~eV, $\langle \Gamma \rangle=2$, $E_{\mathrm{cut}}=260$~keV, $A_{\mathrm{Fe}}=0.65$). The $2$--$10$~keV
number counts predicted by both $2$--$10$~keV LFs are
shown in Fig.~\ref{fig:2to10counts}.}
\label{fig:chisqs}
\end{figure*}

Turning to the six spectral models that incorporate distant and (for
four of them) disc reflection, the best fit
($\chi^2/$dof$=80.5/48=1.68$) to the four LFs is found
for the model with a constant narrow \fe\ EW$=70$~eV (i.e., no Baldwin
effect), and weak (broad \fe\ EW$=54$~eV) disc reflection. The best
fit parameters are $\langle \Gamma \rangle = 1.85 \pm 0.15$,
$E_{\mathrm{cut}}=270^{+170}_{-80}$~keV, and
$A_{\mathrm{Fe}}=0.3^{+0.3}_{-0.15}$ (90\% C.L.). The fit to the LFs
is shown as the solid lines in Figure~\ref{fig:chisqs}.
The presence of the disc reflection component is very marginal, and
removing it increases $\chi^2$ by 0.5, with the other
parameters changing within the error-bars. The presence of the Baldwin
effect, however, gives a $\Delta \chi^2=+9.4$, indicating that
strong reflection is preferred over a wide range of AGN luminosities
(see Sect.~\ref{sub:distant}).

The reduced $\chi^2$ of this fit ($1.68$) is relatively high and is,
strictly speaking, not an acceptable fit to the data. This is either
indicating that the spectral model is not correct or that some of the
error-bars in the LFs are underestimated. The $\chi^2$ of the model
when fit to the individual LFs are indicated in the panels of
Fig.~\ref{fig:chisqs}, and shows that the largest contributions
to the joint $\chi^2$ are from the $0.5$--$2$~keV and $2$--$10$~keV
LFs. Many of the points that make up the $0.5$--$2$~keV LF have small
error-bars and show slight offsets from the power-law shape of the
reference LF. Changes to how $f_2$ is calculated (e.g., using eq.~5
of \citealt{burlon11}) or the $N_{\mathrm{H}}$ distribution makes a negligible difference to $\chi^2$ in
this band. Any additional spectral components such as a warm absorber or a soft excess would be unable to account for these
wiggles in the LF without significant fine-tuning. Therefore, the large
$\chi^2$ found in the $0.5$--$2$~keV band is a result of small
error-bars and peculiar shape of this LF. In contrast, the relatively
high $\chi^2$ in the $2$--$10$~keV band is simply a result of the fact
that the shape of the observed LFs is incompatible with the other
three bands. To illustrate this, the dashed lines in
Fig.~\ref{fig:chisqs} show the predicted LFs when only the
$2$--$10$~keV LF was used to constrain the model. While the fit to the
$2$--$10$~keV LF is excellent (reduced $\chi^2=0.67$), the joint
reduced $\chi^2=33$. The implications of the new $2$--$10$~keV LF fit
is discussed in Sect.~\ref{sub:lf}.

\begin{figure*}
\centerline{
\includegraphics[width=0.5\textwidth]{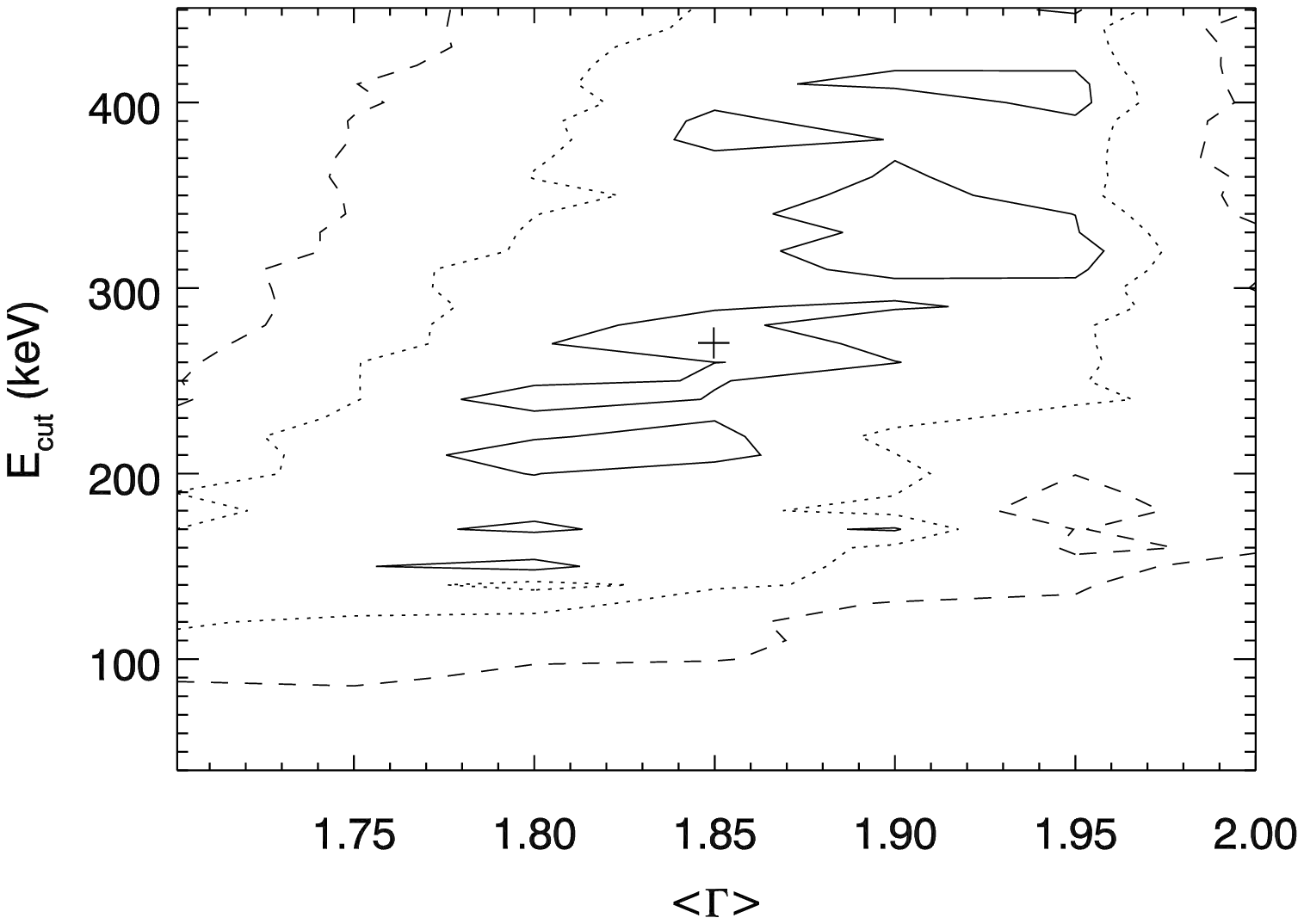}
\includegraphics[width=0.5\textwidth]{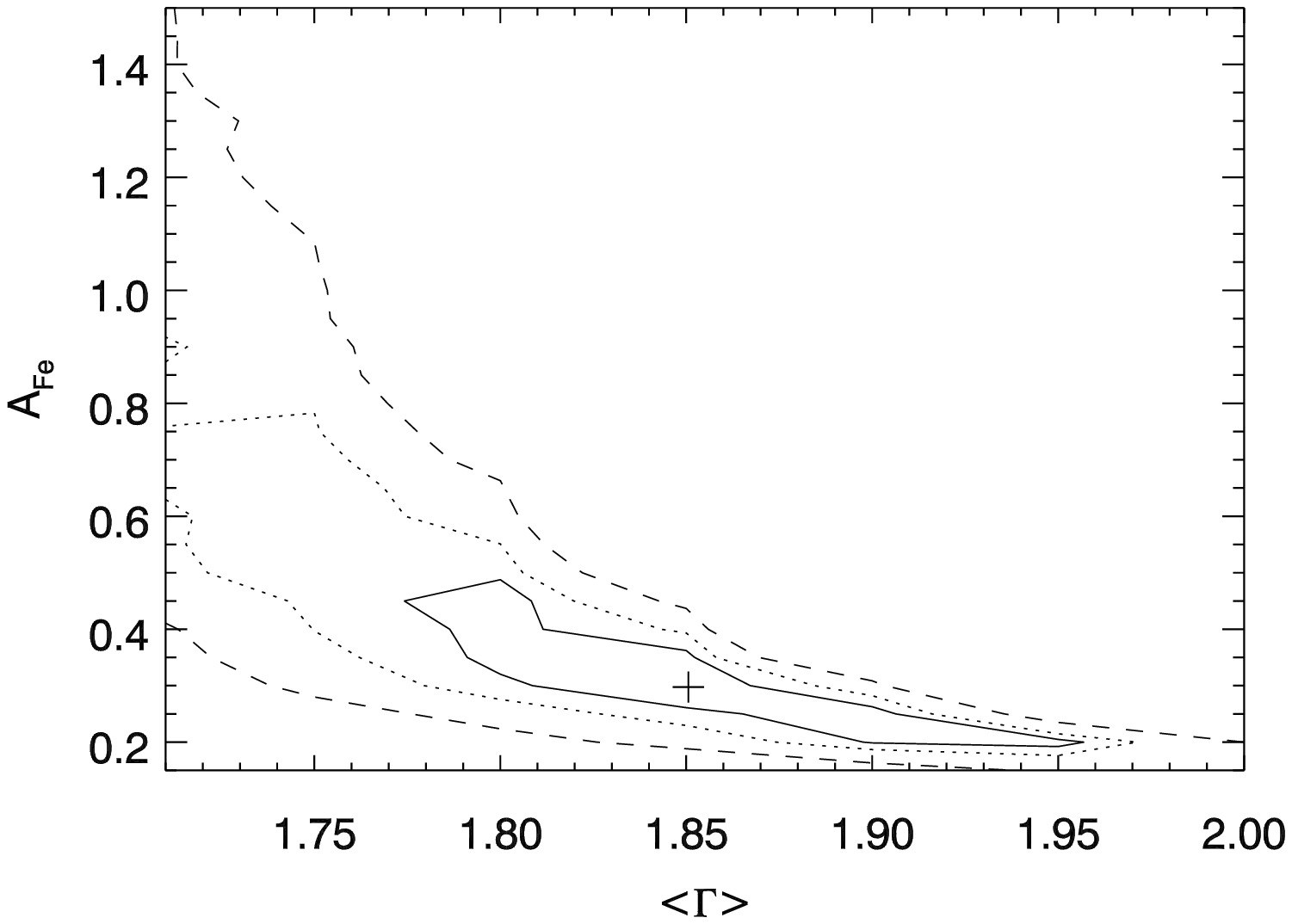}
}
\caption{The 68\% (solid), 90\% (dotted) and 95\% (dashed) confidence
  contours (computed for two parameters of interest) on the spectral
  model that gives the best joint fit to the multi-band LFs
  (Fig.~\ref{fig:chisqs}). The contours are plotted in the
  $E_{\mathrm{cut}}$--$\langle \Gamma \rangle$ (left panel) and
  $A_{\mathrm{Fe}}$--$\langle \Gamma \rangle$ (right panel)
  planes. The plus signs indicate the location of the best-fit
  parameters ($\langle \Gamma \rangle=1.85$, $A_{\mathrm{Fe}}=0.3$ and
  $E_{\mathrm{cut}}=270$~keV).}
\label{fig:contours}
\end{figure*}

Returning to the best fit spectral parameters,
Figure~\ref{fig:contours} plots the 68\%, 90\% and 95\% confidence
contours (computed for two parameters of interest) in the $E_{\mathrm{cut}}$-$\langle \Gamma \rangle$ and
$A_{\mathrm{Fe}}$-$\langle \Gamma \rangle$ planes.
The contours indicate that the tightest constraints are on
$A_{\mathrm{Fe}}$ which is constrained to be sub-solar at the 90\% C.L.
for any value of $\langle \Gamma \rangle$. As the narrow \fe\ EW was
fixed at $70$~eV, this is indicating the presence of strong reflection
in the average spectrum. As $\langle \Gamma \rangle$ becomes harder,
than the reflection can weaken, and the contours shift upward in the
plot. In contrast, the limits on $E_{\mathrm{cut}}$ are
relatively wide, which is not surprising given that the
$14$--$195$~keV band provides only weak constraints on this parameter.

It is interesting to consider what information in the LFs drives the
fit of the spectral shape. To answer this question, all the calculations
were repeated omitting the $0.5$--$2$~keV LF from the $\chi^2$ fitting
and then repeated again with the $3$--$20$~keV LF omitted and the
$0.5$--$2$~keV LF reinstated. As these datasets have the smallest
error-bars they carry the most power in constraining the model. We find
that the $0.5$--$2$~keV LF is crucial to determining $\langle
\Gamma \rangle$. This energy band carries little to no information on
the reflection strength or the high-energy cutoff, but strongly
disfavours $\langle \Gamma \rangle > 1.95$
(Fig.~\ref{fig:contours}). Not surprisingly, the $3$--$20$~keV LF is
most important in constraining the reflection strength as measured by
$A_{\mathrm{Fe}}$. Once this LF was included, the limits on
$A_{\mathrm{Fe}}$ improved significantly, which, as seen in
Fig.~\ref{fig:contours}, affects the values of $\langle \Gamma
\rangle$ and $E_{\mathrm{cut}}$.

\section{Discussion}
\label{sect:discuss}

\subsection{The average spectrum of local AGNs}
\label{sub:avgspect}
\begin{figure}
\centerline{
\includegraphics[width=0.4\textwidth,angle=-90]{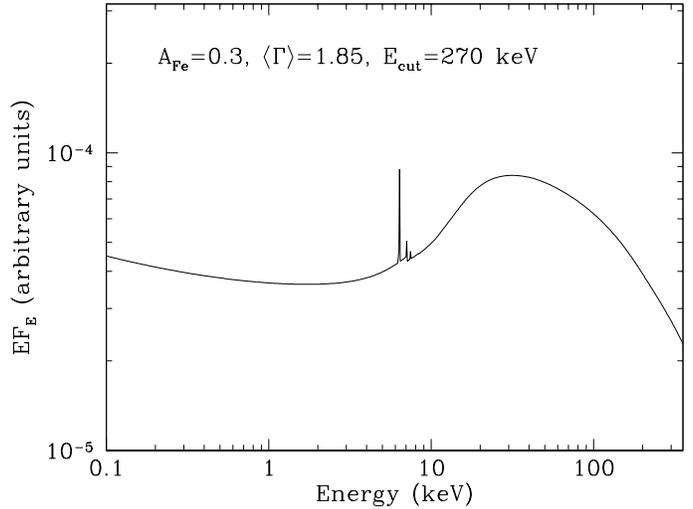}
}
\caption{The average X-ray spectrum of local AGNs as found from
  fitting the local multi-band X-ray LFs (Fig.~\ref{fig:chisqs}). The
  spectral model consists of a Gaussian distribution of power-laws
  with mean photon index $\langle \Gamma \rangle$, dispersion
  $\sigma_{\Gamma}=0.3$, and high energy cutoff
  $E_{\mathrm{cut}}$. A reflection spectrum with the same $\Gamma$
  distribution and $E_{\mathrm{cut}}$ was calculated from
  angle-averaged `pexmon' models \citep{nan07} and added to the primary power-law so
  that the \fe\ EW$=70$~eV \citep{shu10}. Relativistically blurred disc
  reflection is added so that the broad \fe\ EW$=54$~eV \citep{ball10}. }
\label{fig:spectrum}
\end{figure}

The result of our experiment is shown in Figure~\ref{fig:spectrum} --
an average AGN spectrum that is entirely consistent with the $z
\approx 0$ LFs spanning from $0.5$ to $195$~keV.
The derived $\langle \Gamma \rangle=1.85\pm 0.15$ is consistent with
many previous measurements of the mean spectral index from a variety
of recent samples \citep[e.g.,][]{shem06,mol09,burlon11,bn11,ricci11,rmr13,vas13,vmg13},
but is marginally inconsistent with some others \citep{scott11,derosa12,mol13}. The inconsistent values are found from analyses that did not
include reflection, were dealing with non-simultaneous observations, or
had low count rate data. 

Previous constraints on the mean value of $E_{\mathrm{cut}}$ are poor
with values ranging from $\sim 110$~keV to $\sim 300$~keV \citep[e.g.,][]{mol09,ricci11}. The
value derived here ($270^{+170}_{-80}$~keV) is largely consistent with
these measurements. Assuming a corona with $\tau \la 1$, this value
implies that the average AGN corona temperature is $kT \approx
135^{+85}_{-40}$~keV, consistent with prior measurements \citep{zdz00,mol09}.

The third parameter determined by our procedure is the relative
abundance of Fe in the distant reflector,
$A_{\mathrm{Fe}}=0.3^{+0.3}_{-0.15}$. The implication of this value
for models of the distant reflector is discussed below
(Sect.~\ref{sub:distant}), but the reason $A_{\mathrm{Fe}}$ is driven
to low values is to produce a strong Compton reflection hump at high
energies (recall that our procedure fixes the \fe\ EW). Previous
measurements of the reflection strength all involve the $R$
parameter. To compare with the previous results, the fitting procedure
was re-run with $R$ as a variable and $A_{\mathrm{Fe}}=1$ (the
accretion disc reflector was omitted for this run). A good fit was
obtained (reduced $\chi^2=1.63$; $\Delta \chi^2=-2.08$ compared
to the model with variable $A_{\mathrm{Fe}}$) with $R=1.7^{+1.7}_{-0.85}$ (90\%
C.L.), indicating that, indeed, the average reflection strength seems
to be large (the values of $\langle \Gamma \rangle$ and
$E_{\mathrm{cut}}$ obtained in this fit are $1.85$ and
$220$~keV). This value of $R$ is consistent with earlier measurements
(e.g., \citealt{mol09,burlon11,vas13,vmg13}, but see also \citealt{rmr13}). 

\begin{figure}
\centerline{
\includegraphics[width=0.4\textwidth,angle=-90]{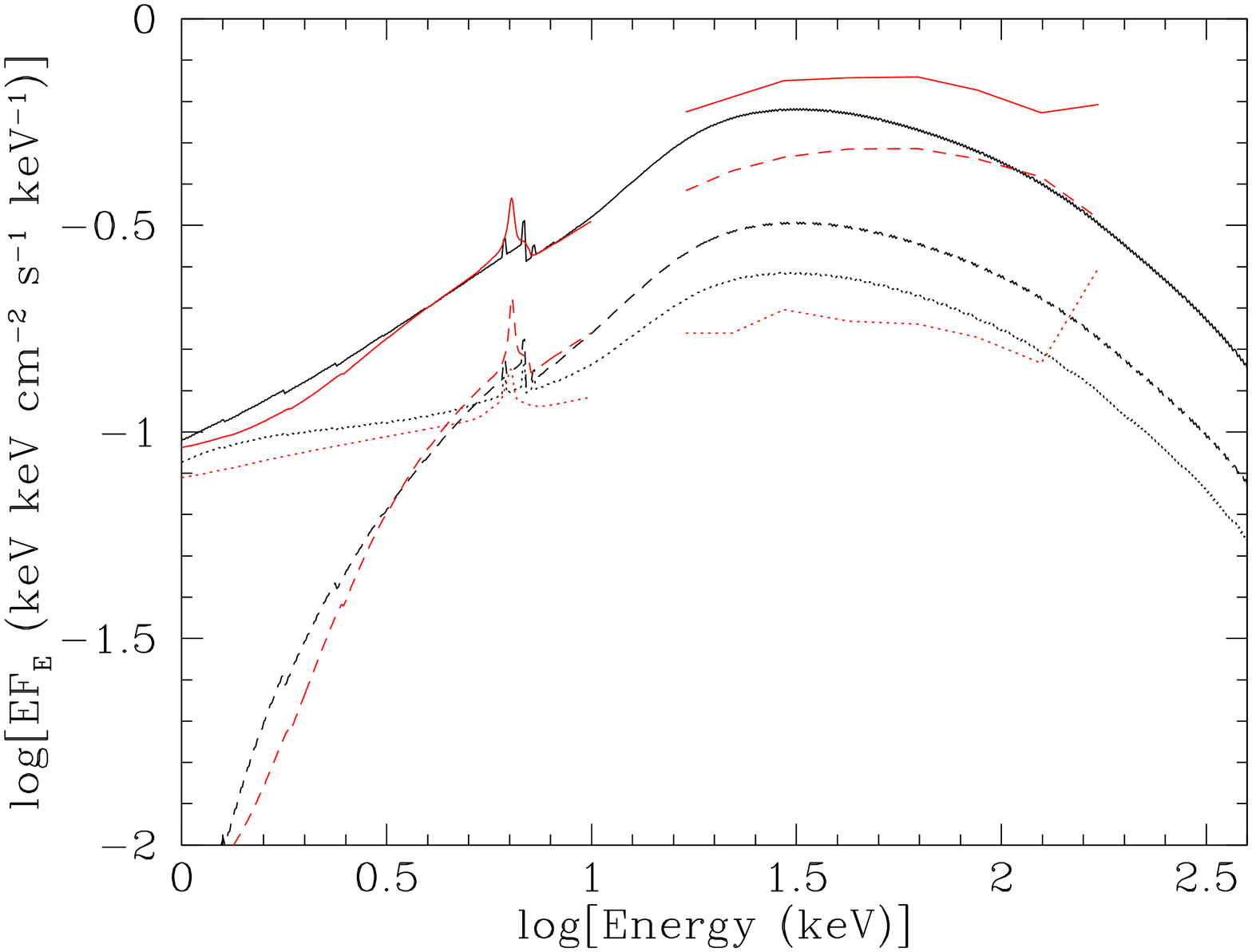}
}
\caption{The red lines plot the stacked spectra of 96 \bat\ selected
local AGNs \citep{vmg13} (solid=all AGNs; dashed=AGNs with $22 < \log
N_{\mathrm{H}} < 24$; dotted=AGNs with $\log N_{\mathrm{H}} <
22$). The black lines shows the result of summing the derived local
AGNs average spectrum (Fig.~\ref{fig:spectrum}) over the
\citet{ajello12} LF while including the absorption distribution. To
better compare the shapes of the two spectra, the models were scaled
so that the total intensity (solid lines) were equal at $\approx
5$~keV. The $2$--$10$~keV slope of the predicted spectrum is
$\Gamma=1.44$, very similar to the observed slope. The contributions
from unobscured and Compton-thin obscured AGNs are in good agreement;
however, the stacked spectra show stronger reflection in obscured AGNs
and moderately weaker reflection in the unobscured AGNs.}
\label{fig:vas}
\end{figure}

\citet{vmg13} recently presented the stacked spectrum of a complete
sample of 95 \bat-selected local AGNs in the Northern Galactic
Cap. As the individual spectra were not corrected for absorption, the
resulting sum had the characteristic $2$--$10$~keV $\Gamma \approx 1.4$
power-law of the XRB spectrum. If the average $z \approx 0$ AGN
spectrum of Fig.~\ref{fig:spectrum} is accurate, then summing this
spectrum over the \citet{ajello12} LF [while including the \citet{burlon11}
$N_{\mathrm{H}}$ distribution and the $f_2$--$L$ relation
(Fig.~\ref{fig:f2vsL})] should yield a similar
result. Figure~\ref{fig:vas} shows the result of this experiment where
the model was scaled so that the two integrated spectra (solid lines; the \citet{vmg13} result is
in red) are normalized to be the same at an energy of $\approx 5$~keV.
In general, the two spectral shapes are in good agreement (the
predicted spectrum has $\Gamma=1.44$ in the $2$--$10$~keV band), with
the stacked \bat\ sources showing a more pronounced hard X-ray
hump. The reason for this is found by examining the dotted and dashed
lines which plot the contributions from unobscured ($\log
N_{\mathrm{H}} < 22$) and Compton-thin obscured sources (i.e., $22 <
\log N_{\mathrm{H}} < 24$). Here, one can see that the stacked
Compton-thin AGNs show a stronger reflection hump than predicted while the unobscured
AGNs present a slightly weaker hump, however, the measured $R$ values 
agree with the model one within the errors \citep{vmg13}. Interestingly, the contributions from
obscured and unobscured AGNs agree very well (these lines were not
individually adjusted -- all three black curves were moved by the same
factor to normalize the total) indicating that the Northern Galactic
Cap sources well sample the \citet{burlon11} $N_{\mathrm{H}}$
distribution. All in all, the agreement of the two spectral shapes
seen in Fig.~\ref{fig:vas} supports the accuracy of our derived
spectrum and the effectiveness of the LF-fitting procedure.

\subsubsection{Implications for fitting the XRB spectrum}
\label{subsub:xrb}
The value of $\langle \Gamma \rangle$ measured here is consistent with the typical value of $1.9$ used by XRB
models \citep[e.g.,][]{gch07,tre09,ball11}. Similarly, the derived $E_{\mathrm{cut}}$ is also close to the values
($200$--$300$~keV) used in XRB synthesis models.  In contrast, there
is more variation on the assumption made for the reflection strength:
\citet{gch07} and \citet{ball11} assume $R \approx 1$ and $A_{\mathrm{Fe}}=1$, but with the reflection strength dropping off with
luminosity; alternatively, \citet{tre09} use $R=1$ and
$A_{\mathrm{Fe}}=2$ at all $L$. This last spectrum has a strong
reflection hump and is most similar to the average spectrum measured
here. When fitting the XRB spectrum, one must also deal with
uncertainties in the evolution of the X-ray LF \citep{db09}, the Compton-thick
fraction (and its possible evolution; \citealt{db10}), and the
absorption distribution. These factors, combined with the degeneracies
involved in fitting the spectrum \citep{ay12}, means the XRB spectrum
can still be fit with an average spectral shape within the
measured uncertainties we derived for $\langle \Gamma \rangle$,
$E_{\mathrm{cut}}$ and $A_{\mathrm{Fe}}$.

\subsection{The 2--10 keV AGN luminosity function}
\label{sub:lf}
As indicated above, the best multi-band fit to the LFs resulted in a
$2$--$10$~keV LF that is not a good representation of the ones
measured by \heao\ and \maxi. Indeed, as seen in
Figure~\ref{fig:chisqs}, a good fit to the 2--10~keV LF data results
in a very poor fit to the $0.5$--$2$~keV and $3$--$20$~keV
LFs. Therefore, it appears that the measured local AGN 2--10~keV LF is
incompatible with the shape of the LFs determined by \bat, \rxte\ and \rosat.

\begin{figure}
\centerline{
\includegraphics[width=0.5\textwidth]{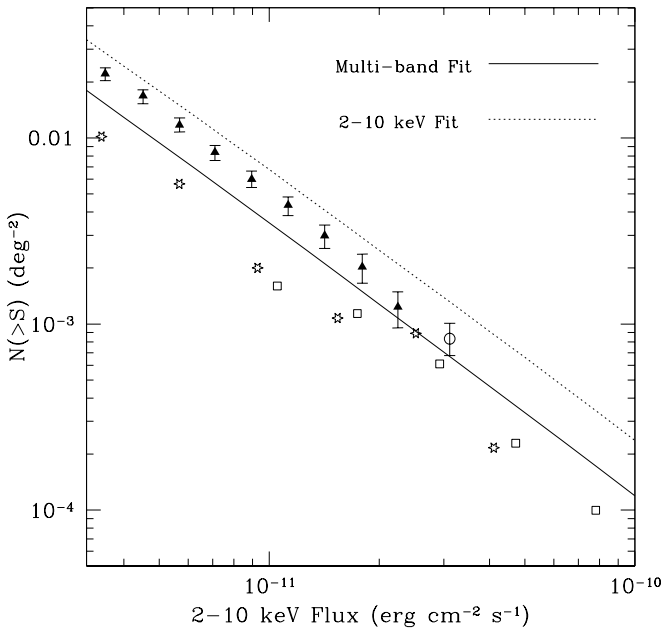}
}
\caption{The solid line plots the 2--10~keV number count predicted by
  the AGN spectral model derived from the multi-band LF fit, while the
dotted line shows the predictions using a spectral model that best
fits the local $2$--$10$~keV LF (Fig.~\ref{fig:chisqs}). The data points
are taken from an extragalactic sample from the \xmm\ Slew Survey
(solid triangles; $\approx 80$\% AGNs; \citealt{war12}), the \heao\ AGN sample (open circle;
\citealt{pic82}), and follow-up surveys of \bat\ selected
sources (open squares and stars; \citealt{winter09} and
\citealt{vas13}, respectively).}
\label{fig:2to10counts}
\end{figure}

\begin{figure}
\centerline{
\includegraphics[width=0.5\textwidth]{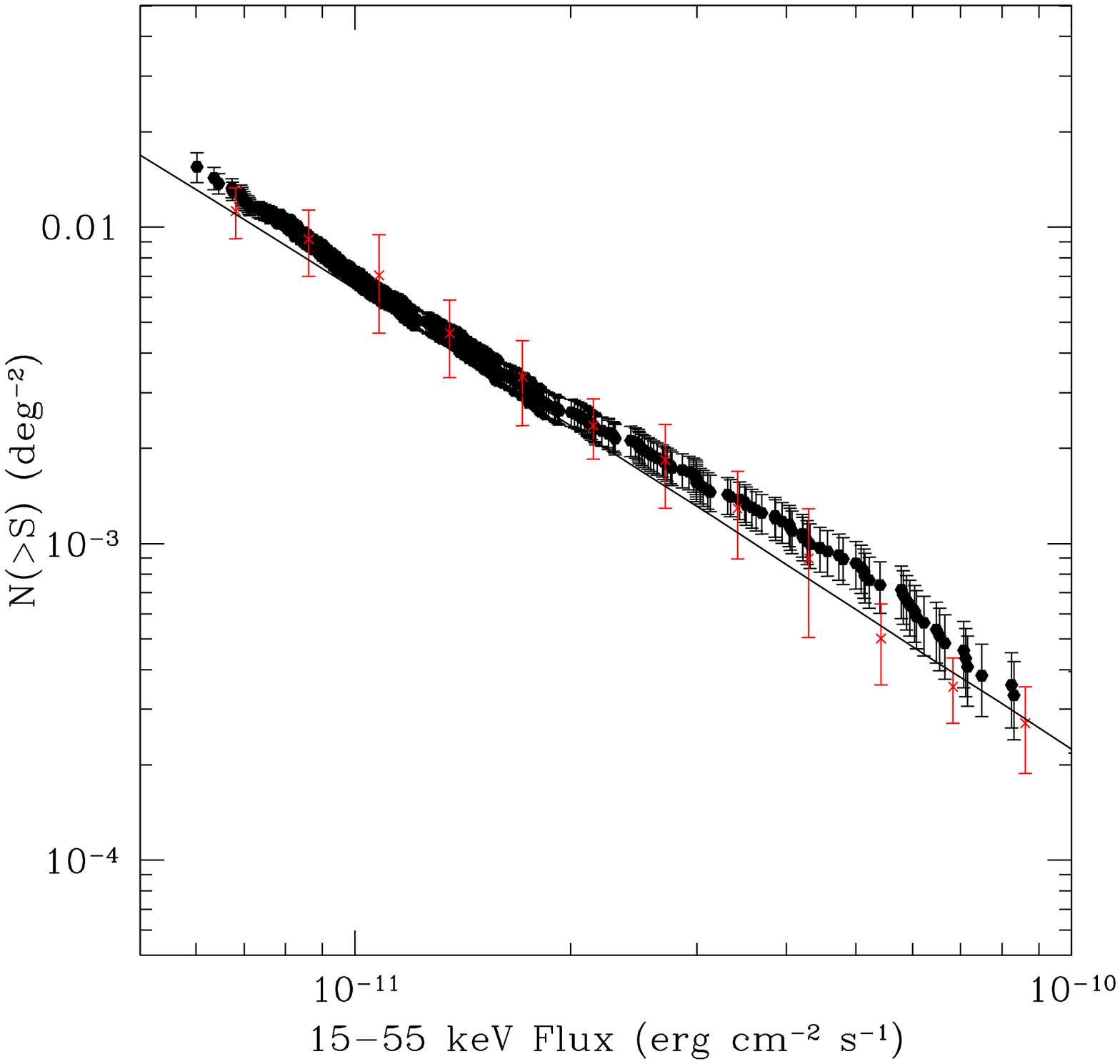}
}
\caption{The solid line plots the $15$--$55$~keV number count predicted by
  the AGN spectral model derived from the multi-band LF fit (Fig.~\ref{fig:chisqs}). The data points
are taken from the 60 month \bat\ survey (black points;
\citealt{ajello12}), the \integral\ measurements by \citet{kriv10}
(red crosses; converted to the $15$--$55$~keV band by
\citealt{ajello12}). Although we use the $15$--$55$~keV LF described by
\citet{ajello12}, the number counts are slightly underpredicted due to
omitting the evolution of the LF. Therefore, the predicted
$2$--$10$~keV LF shown in figure~\ref{fig:2to10counts} will also be
slightly below its true value.}
\label{fig:15to55counts}
\end{figure}

As this is a surprising result, the $2$--$10$~keV AGN number counts
were calculated using the two predicted LFs and spectral models. As
these are local LFs, the number counts were only computed for
relative bright fluxes. Figure~\ref{fig:2to10counts} compares the
predicted counts to measurements from \heao\ \citep{pic82}, the \xmm\ Slew
Survey (an extragalactic sample that is $\approx
20$--$25$\% galaxies; \citealt*{war12}), and follow-up surveys of \bat\ sources \citep{winter09,vas13}. 
Apart from the \heao\ point, all these data are entirely independent
from those used in the LF fitting and therefore provides a test of the
veracity of the predicted LFs. Fig.~\ref{fig:2to10counts} shows that
the LF found from fitting only the $2$--$10$~keV band overpredicts the
measured number counts at all fluxes. In contrast, the LF determined
from the multi-band fit underpredicts the counts from the \xmm\ Slew
Survey. To determine if an overprediction or underprediction is more
likely, we computed the $15$--$55$~keV counts from the adopted
$15$--$55$~keV LF (Fig.~\ref{fig:15to55counts}). As the data that
comprise the observed $15$--$55$~keV counts were used to measure the
LF, the predicted number counts should closely match the observations.
Figure~\ref{fig:15to55counts} shows that the model number counts
slightly underpredict the observations, most likely a result of
omitting any evolution of the LF and halting the integration at
$z=0.2$. 
Therefore, the predicted $2$--$10$~keV number counts will also be slightly
underestimated, which indicates that the model from the multi-band LF
fit best describes the observed number counts in the $2$--$10$~keV
band. The level of correction implied by Figure~\ref{fig:15to55counts}
will not be enough for the model to exactly match the \xmm\ Slew
Survey data, but these data include a non-negligible fraction of
galaxies, and the nature of the slew survey made identification of
counterparts difficult at low fluxes \citep{war12}. Therefore, the
measured number counts from this experiment are likely to be slightly
too large, especially at low fluxes. Finally, we note that the
multi-band fit does overpredict many of the datapoints from
\citet{winter09} and \citet{vas13}, but no error-bars are reported for
these measurements, so the degree of mismatch is hard to estimate. 

The number counts support the finding that the observed 2--10 keV LFs
reported by \heao\ and \maxi\ are incompatible with the shape of the
LFs in other energy bands. Fig.~\ref{fig:chisqs} shows that the shapes
disagree both at the high and low luminosity ends. The high luminosity
end is most likely contaminated by AGNs at $z > 0.1$ that were
included in the `local' sample --- the inclusion of these objects
would push up the high luminosity end of the LF due to the luminosity
evolution of the LF at $z \ga 0.1$ \citep[e.g.,][]{ajello12}. It is less clear what
is causing the overprediction at the low-luminosity end, but it may be
related to the small number of AGNs in both samples used to construct
the LF. Recall from Sect.~\ref{sect:calc} that out of the five LFs
used here, the 2--10~keV LF measurements were constructed from samples
of AGN around half the size of those used in the other
bands\footnote{Since the measured \heao\ and \maxi\ $2$--$10$~keV LFs are likely affected
  by the luminosity evolution of the true local LF, we repeated the spectral fitting without the
  $2$--$10$~keV LFs in order to assess the impact of evolution on
  the derived average spectral shape. The resulting best fit spectral
  model was completely consistent with the one shown in
  Fig.~\ref{fig:spectrum}, indicating that LF evolution does not
  impact the derived average AGN spectrum.}.

\begin{figure}
\centerline{
\includegraphics[width=0.5\textwidth]{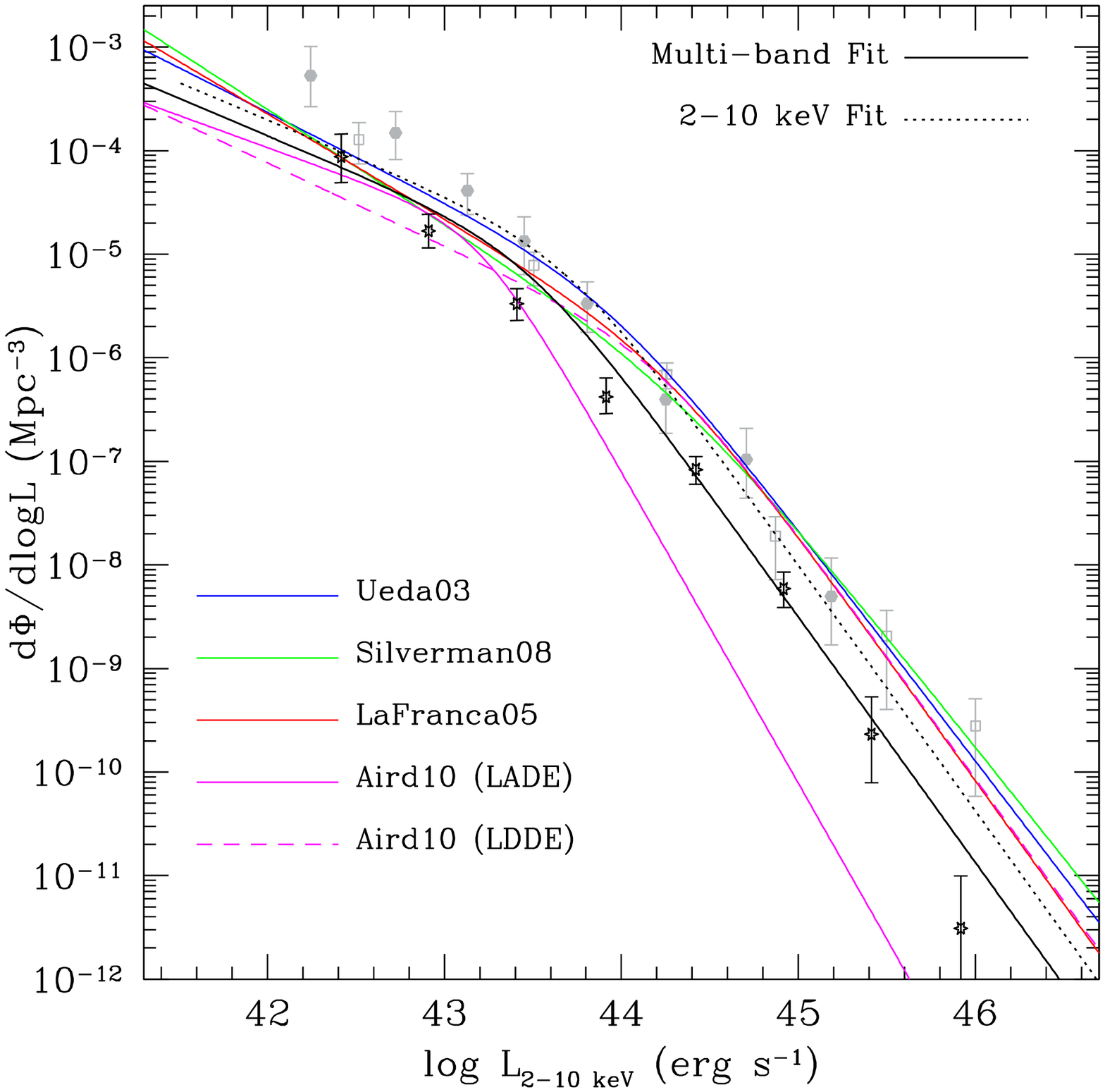}
}
\caption{Comparison of the local ($z \approx 0$) 2--10~keV AGN LF
  determined by the multi-band fit (black solid line) with the LFs published by
  \citet{ueda03}, \citet{laf05}, \citet{silv08} and \citet{aird10}
  (coloured lines). The black stars plot the de-evolved $z=0$ LF
  estimated by \citet{della08} from the \xmm\ Hard Bright Survey. The grey data points are the same as those plotted
  in Fig.~\ref{fig:chisqs}. The dotted black line plots the best
  2--10~keV LF obtained by our procedure when ignoring the fits to the
  other three LFs (see also Fig.~\ref{fig:chisqs}). The 2--10~keV LF obtained by our
  multi-band approach is strikingly different than the other
  measurements except for the \citet{della08} estimate.}
\label{fig:2to10LF}
\end{figure}

A revision to the local 2--10~keV LF will have important implications
in understanding the evolution of AGNs, as it will change the
zero-point for all determinations of the evolving LF. To illustrate
this, Figure~\ref{fig:2to10LF} compares the 2--10~keV LF determined
by the multi-band fitting (thick black line) to the the $z=0$ LFs predicted by several
evolving AGN LFs from the literature (coloured lines). 
For reference, the figure also shows the observed 2--10~keV LF as the
grey data. This plot clearly shows that every other LF (except for the
LADE model of \citealt{aird10}) closely follows the observed 2--10~keV
LF, especially at the high luminosity end, and is therefore
\emph{incompatible with the measured LFs at all other bands}. The lone
exception is the de-evolved $z=0$ LF estimated by \citet{della08} from
the \xmm\ Hard Bright Survey (black stars) which, although dependent
on the model used to de-evolve the LF, is consistent with the
multi-band fit at high and low luminosities. It is
crucial that the evolution of the AGN LF in the 2--10~keV band be
recomputed using the $z=0$ LF reported here (see
Table~\ref{table:LFdata}) as the zero-point. Once this is done XRB synthesis
models will need to be updated in order to revise predictions of the
Compton-thick space density at high-$z$ \citep[e.g.,][]{tre09,ball11,ay12}.

\begin{table}
\caption{The $2$--$10$~keV LF of $z \sim 0$ AGNs as determined by the
  multi-band fit (dark solid line in Fig.~\ref{fig:2to10LF}).}
\label{table:LFdata}
\begin{center}
\begin{tabular}{cc}
$\log L_{\mathrm{2-10\ keV}}$ & $\log(d\Phi/d\log L)$\\
($h_{70}^{-2}$~erg~s$^{-1}$) & ($h_{70}^{3}$~Mpc$^{-3}$)\\ \hline 
41.30 &	 $-$3.35\\
41.56 &	 $-$3.54\\
41.82 &	 $-$3.73\\
42.08 &	 $-$3.91\\
42.34 &	 $-$4.11\\
42.60 &	 $-$4.30\\
42.86 &	 $-$4.51\\
43.12 &	 $-$4.76\\
43.38 &	 $-$5.07\\
43.64 &	 $-$5.47\\
43.90 &	 $-$5.98\\
44.16 &	 $-$6.54\\
44.42 &	 $-$7.14\\
44.68 &	 $-$7.75\\
44.94 &	 $-$8.36\\
45.20 &	 $-$8.98\\
45.46 &	 $-$9.59\\
45.72 &	 $-$10.21\\
45.98 &	 $-$10.82\\
46.24 &	 $-$11.44\\
46.50 &	 $-$12.06\\
46.76 &	 $-$12.67\\
47.02 &	 $-$13.29\\
47.28 &	 $-$13.90\\
47.54 &	 $-$14.52\\
47.80 &	 $-$15.14\\
\end{tabular}
\end{center}
\end{table}

\subsection{Constraints on the distant reflector}
\label{sub:distant}
Our procedure also provided interesting constraints on the distant
reflector. First, the fits preferred that the reflection strength
remain constant as a function of AGN luminosity \citep[e.g.,][]{vmg13}; that is, the presence
of a X-ray Baldwin effect was not required. The strength of any X-ray
Baldwin effect has been the source of debate for several years, as it
requires precise observations over a wide range of luminosities and
being able to control for variability. If present, it might be
connected to the obscuring material around the AGN, as \citet{ricci13} recently
showed that the observed decrease in $f_2$ with luminosity could give
rise to a Baldwin effect of the right slope. The fact that no Baldwin
effect is preferred in the LF fits would indicate that the distant
reflector is not associated with the obscuring material, but arises
from another source that is common to AGNs of all luminosity, such as
the outer regions of the accretion disk or broad-line region clouds
\citep{pet02,nan06,bian08}. Another intriguing possibility is that the distant reflector
is associated with the mid-infrared (12~$\mu$m) emitter that appears to be
independent of the X-ray luminosity and AGN classification
\citep{hor08,gan09,hon10}. Its is also important to remember that both
the evidence for and against the X-ray Baldwin effect is concentrated
at $2$--$10$~keV luminosities $\la 10^{44}$~erg~s$^{-1}$; it is
still unknown precisely how the reflection strength behaves at
larger luminosities.

The average spectrum derived in Sect.~\ref{sect:res} strongly
indicates that the distant reflector has a sub-solar Fe
abundance. Again, the reason the fit was driven to this result was to
simultaneously satisfy the constraints of strong reflection and a
\fe\ EW of 70~eV. Indeed, the best fitting spectral model that used
the $R$ parameter (and found $R=1.7$) predicts a narrow \fe\ EW of
216~eV, much larger than the typical observed values \citep[e.g.,][]{shu10}. However,
a sub-solar Fe abundance in the nuclear environment of AGN host
galaxies is problematic, as there is a wealth of
evidence that metallicity increases to super-solar abundances in the
AGN environment (e.g., \citealt*{nag06} and references therein). Therefore, it is highly unlikely that the abundance
$A_{\mathrm{Fe}}=0.3$ found by this procedure is pointing to the true
iron abundance. Changing the abundance set to the \citet{lodd03} measurement that
has a smaller iron abundance does not alter the derived $A_{\mathrm{Fe}}$. Part of the problem may be the geometric assumptions
built into the `pexmon' model, as the Compton backscattered and
associated line emission are based on an infinitely thick disk
geometry. Repeating this experiment with a neutral reflection model
that can handle a wider range of geometries (e.g., MYTorus; \citealt{yaq12}) would
be able to test this assertion, with the limitation of introducing
additional free parameters. Alternatively, the low value of
$A_{\mathrm{Fe}}$ may indicate that the reflecting medium is
dusty. Gas-phase iron is heavily depleted onto grains in dusty gas,
but the iron in the grains can still produce \fe\ emission
\citep[e.g.,][]{fer13}. In this scenario, the EW of the line would depend on the
details of the gas-grain mixture as well as geometry, so further
investigation of this idea will be the subject of future
work. However, this explanation for the value of $A_{\mathrm{Fe}}$
would support the association of the distant reflector with the
12~$\mu$m emitter.

\section{Conclusions}
\label{sect:concl}
This paper presents a measurement of the properties of the average $z\approx 0$ AGN
spectrum between $0.5$ and $200$~keV by simultaneously fitting the
measured AGN LFs in the $0.5$--$2$~keV, $2$--$10$~keV, $3$--$20$~keV
and $14$--$195$~keV bands. The $15$--$55$~keV LF measured by
\bat\ \citep{ajello12} is used as the best determination of the true $z\approx 0$
LF and the spectral shape was varied to find the best fit to the other
four bands. The spectral parameters constrained were the mean photon
index of AGNs, $\langle \Gamma \rangle$, the cutoff energy of the
power-law, $E_{\mathrm{cut}}$, and the strength of the distant
reflector as parameterised by its iron abundance,
$A_{\mathrm{Fe}}$. The luminosity dependence of the LFs also allowed a
test for the presence of the X-ray Baldwin effect. The principle
findings of this study are:
\begin{itemize}
\item The best fitting mean AGN spectral model has $\langle \Gamma \rangle = 1.85 \pm 0.15$,
$E_{\mathrm{cut}}=270^{+170}_{-80}$~keV, and
  $A_{\mathrm{Fe}}=0.3^{+0.3}_{-0.15}$ (90\% C.L.). The low iron
  abundance indicates the need for strong reflection in the average
  spectrum given the assumed \fe\ EW. The reflection strength is
  equivalent to $R=1.7^{+1.7}_{-0.85}$ (90\% C.L.). The absence of the
  X-ray Baldwin effect is mildly preferred by the fit ($\Delta
  \chi^2=+9.4$ when the effect is included). These values are
  consistent with most previous measurements of the average AGN
  spectrum.

\item The shape of the local $2$--$10$~keV AGN LF as measured by \heao\ and
  \maxi\ is incompatible with LFs measured by \bat\ and other
  instruments. This is likely due to the inclusion of higher $z$
  sources in the sample and the small size of the samples. As a
  result, the evolving $2$--$10$~keV AGN LFs used in XRB modeling need
  to be revised. The $2$--$10$~keV LF predicted by our procedure is
  provided in Table~\ref{table:LFdata} and is consistent with
  the LFs measured in the other bands. This new LF should be used as
  the zero-point for determining the evolution of the AGN LF in the
  $2$--$10$~keV band.

\item The procedure indicates that strong distant reflection is
  preferred in local AGNs of all luminosities. That is, we found no
  evidence of a X-ray Baldwin effect. This result implies that the
  distant reflector is not associated with the AGN obscuration zone,
  which does evolve strongly with luminosity. The sub-solar iron
  abundance may be a result of the assumed disc geometry employed by
  `pexmon', but may also be indicating that the \fe\ line is emitted
  by dust grains. If this is the case then the distant reflector must
  lie outside the dust sublimation zone and may
  be plausibly associated with the 12~$\mu$m emitter that is observed
  to be correlated with the X-ray luminosity of all AGNs.  

\end{itemize}

\section*{Acknowledgments}
This work was supported in part by NSF award AST 1008067 to DRB. The
author thanks T.\ Kallman, R.\ Mushotzky and M.\ Ajello for helpful
discussions. M.\ Ajello and R.\ Vasudevan are also acknowledged for
sending published data in an electronic format.

%%%%%%%%%%%%%%%%%%%%%%%%%%%%%%%%%%%%

\bsp % ``This paper has been produced using the ...''

\label{lastpage}

\end{document}